\documentclass[twocolumn]{aastex6}
\usepackage{amsmath,amssymb,amsfonts}
\usepackage{graphicx}
\usepackage{lipsum}
\usepackage{natbib}
\usepackage{color,soul}
\usepackage{hyperref}
\usepackage{marginnote}

\newcommand{\sw}[1]{\texttt{#1}}
\newcommand{\sqd}{sq.~deg.}

\newcommand{\IUCAA}{Inter-University Centre for Astronomy and 
  Astrophysics, Post Bag 4, Ganeshkhind, Pune 411 007, India}
\newcommand{\IISER}{Indian Institute of Science Education and Research Pune, Dr. Homi Bhabha Road, Pashan,  Pune  411008, India}
\newcommand{\WSU}{Department of Physics \& Astronomy, Washington State University,
1245 Webster, Pullman, WA 99164-2814, U.S.A}

\shorttitle{Geography and EM followup of GW sources}
\shortauthors{Srivastava et al.}
\begin{document}


\title{Geographic and seasonal influences on optical followup of gravitational wave events}

\author{Varun Srivastava\altaffilmark{1,2}, Varun Bhalerao\altaffilmark{3}, Aravind P. Ravi\altaffilmark{4,5}, Archisman Ghosh\altaffilmark{5,6}, Sukanta Bose\altaffilmark{3,7}}
\altaffiltext{1}{\IISER}
\altaffiltext{2}{\href{mailto:varun.srivastava@students.iiserpune.ac.in}{varun.srivastava@students.iiserpune.ac.in}}
\altaffiltext{3}{\IUCAA}
\altaffiltext{4}{Indian Institute of Science Education and Research Kolkata, Mohanpur, West Bengal 741252, India}
\altaffiltext{5}{International Centre for Theoretical Sciences, Tata Institute of Fundamental Research, Survey~No.~151, Shivakote, Hesaraghatta Hobli, Bengaluru North 560089, India}
\altaffiltext{6}{Nikhef -- National Institute for Subatomic Physics, Science Park~105, 1098~XG Amsterdam, The~Netherlands}
\altaffiltext{7}{\WSU}

\begin{abstract}
We investigate the effects of observatory locations on the probability of discovering optical/infrared counterparts of gravitational wave sources. We show that for the LIGO--Virgo network, the odds of discovering optical/infrared (OIR) counterparts show some latitude dependence, but weak or no longitudinal dependence. A stronger effect is seen to arise from the timing of LIGO/Virgo observing runs, with northern OIR observatories having better chances of finding the counterparts in northern winters. Assuming identical technical capabilities, the tentative mid-2017 three-detector network observing favors southern OIR observatories for discovery of EM counterparts.
\end{abstract}


\section{Introduction}\label{sec:intro}
The detection of gravitational waves (GW) by LIGO \citep{gw150914,lsc16} marks the beginning of the era of gravitational wave astronomy. Continued improvements in the sensitivity of GW detectors will increase the frequency of detections, enabling detailed study of a variety of sources and source populations.

An key step in the study of GW sources is the detection of electromagnetic (EM) counterparts. While GW signals carry information about physical and geometric properties of the source, study of EM counterparts will yield complementary information necessary to complete our astrophysical understanding of the source~\citep{nkg12,spf+14}. Several groups around the world partook in efforts to follow-up the first gravitational wave detections \citep{aaa+16,aaa+16b}, leading to the discovery of a candidate gamma ray signal potentially associated with binary black hole merger event GW150914~\citep{cbg+16}. There is greater potential for the existence of EM counterparts of GW sources like binary neutron star mergers or supernovae, and several groups will take part in followup activities for future gravitational wave triggers\footnote{A partial list of groups which have signed memoranda of understanding with the LIGO--Virgo collaboration is available at \url{https://gw-astronomy.org/wiki/LV_EM/PublicParticipatingGroups}.}.  

The localization of a GW source by a pair of gravitational wave detectors depends on source parameters and signal strength, and is rather coarse: the 90\% credible regions of the sky localization often span hundreds of square degrees~\citep{spf+14,lsc16}. 
Imaging such large sky areas to find specific transient counterparts poses formidable challenges~\citep[cf.][]{skc+15}. Various aspects of this challenge have been examined in detail, including theoretical modeling of light curves of GW events \citep[etc]{th13,kfm15}; assessing the detectability of EM counterparts \citep{mb12,cb15}; comparing followup capabilities of various facilities \citep{nkg12,kn14}; and strategies for coordination and optimal followup~\citep{sps12,chm+15,gbn+15,rsg+16}. The search for optical counterparts of GW sources is among the driving scientific interests for upcoming projects like  BlackGEM\footnote{BlackGEM --- \url{https://astro.ru.nl/blackgem/}}~\citep{bgn+15} and the Gravitational-wave Optical Transient Observer (GOTO\footnote{GOTO --- \url{http://www.goto-observatory.org/}}).

\begin{figure*}[!thbp]
\begin{center}
\includegraphics[width=\textwidth]{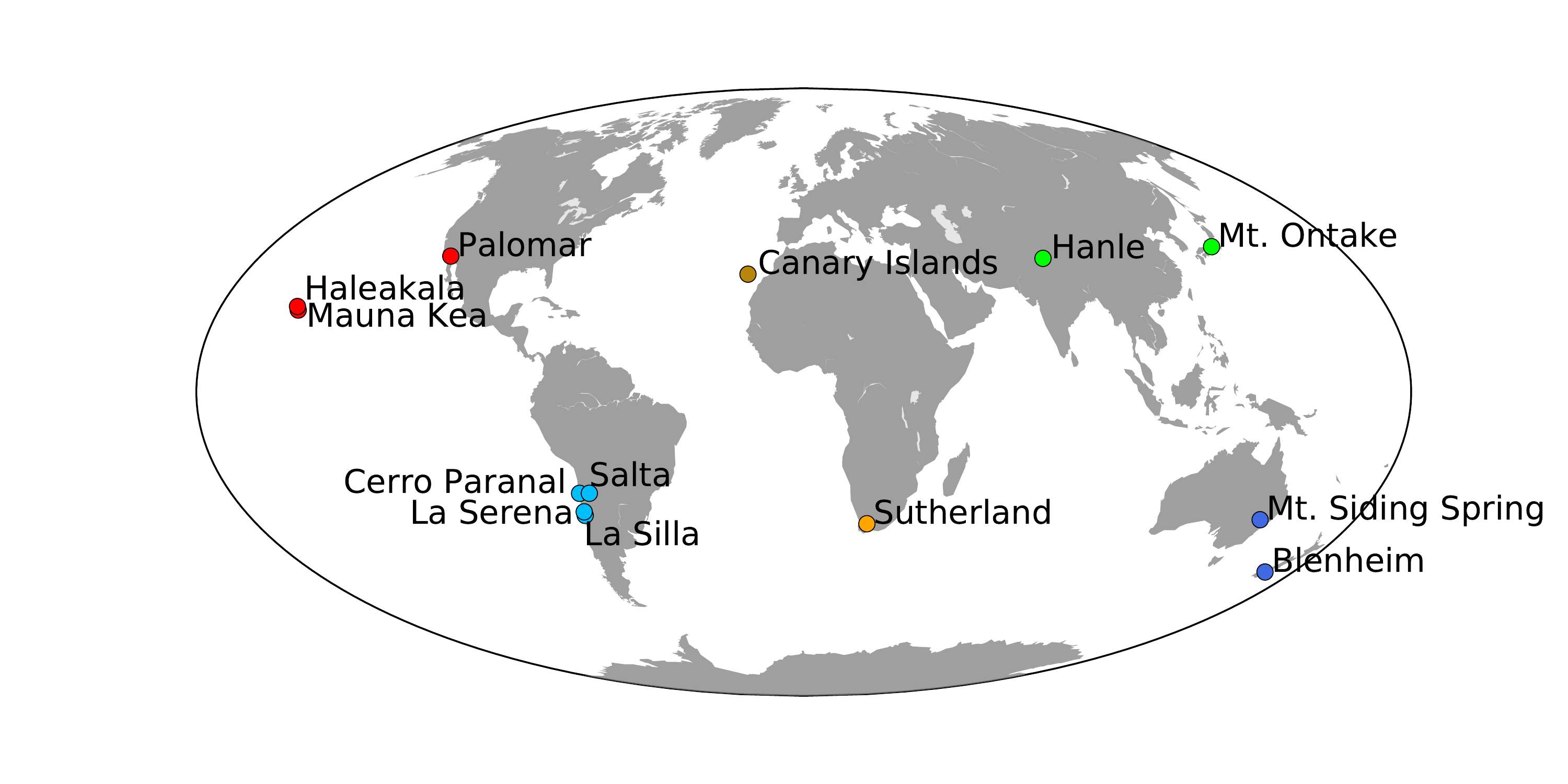}
\caption{Locations of observatories that are considered in this work. We include all ground-based 
optical/infrared observatories that were involved in following up GW150914. We also include Hanle as a representative observatory in Asia.}
\label{fig:obsloc}
\end{center}
\end{figure*}

In this work, we examine another factor that can influence the odds of successful followup: the location of the observatory. GW detectors are not uniformly sensitive to the entire sky, which introduces a sky position dependent bias in the detection and localization of sources~\citep{fairhurst11}.
It is then fair to ask, for instance, if observatories located on the same continent as the two LIGO detectors are ``better placed'' for the search for electromagnetic counterparts of gravitational wave sources. 
A second location-related effect comes from the timing of the LIGO science runs. For instance, for an observing run during northern winter would mean that on an average, a larger part of the localization region is visible to northern observatories during the long nights. It is our aim to examine the effects of these two factors on the follow-up capabilities of various observatories~\citep[cf.][]{cev+16}.

We frame our question and describe our methods in \S\ref{sec:method}. We explore the effects of seasons on observing capabilities of various telescopes in \S\ref{sec:SESE}. We consider the overall effects of geographic location and timing of the LIGO observing runs for two- and three- gravitational wave detector networks in \S\ref{sec:fullsim}, and conclude by discussing the implications in \S\ref{sec:discussion}.

\section{Method}\label{sec:method}
The principal aim of this work is to investigate the effects of (1) location, and (2) seasons on the probability of finding EM counterparts of GW sources from ground-based observatories.

To undertake these comparisons, we need to disentangle the telescope capabilities from location and seasonal effects. We can phrase the question as follows: ``If a telescope based at site $X$ can cover $N$~\sqd\ on the sky to the requisite sensitivity in a single night, what would be the probability of it finding the EM counterpart?'' We answer this question by simulating follow-up optical/infrared observations of a set fake gravitational wave events from various ground-based locations at various times.

As a representative sample of locations of ground-based observatories, we select all optical/infrared telescopes that participated in the followup of GW150914~\citep{aaa+16b}. To fill a hole in Asia, we include Hanle~\citep{prabhu06}, the site of the upcoming 0.7~m robotic telescope, the Indian element of  the ``Global Relay of Observatories Watching Transients Happen'' (GROWTH\footnote{GROWTH --- \url{http://growth.caltech.edu}.}). The sites considered in this work are shown in Figure~\ref{fig:obsloc}, and listed in Table~\ref{tab:obsloc}.

\floattable
\begin{deluxetable}{llcccll}
\tabletypesize{\scriptsize}
\tablewidth{\textwidth}
\rotate
\tablecaption{Locations of ground-based Optical/IR observatories that followed up GW150914\label{tab:obsloc}}
\tablehead{\colhead{Site} & \colhead{Telescope/Instrument} & \colhead{Latitude} & \colhead{Longitude} & \colhead{Altitude} & \colhead{Reference} & \colhead{URL}}
\startdata
Blenheim, New Zealand & BOOTES-3 &  45\degr S & 169\degr 41\arcmin E & 27 m  &  \citet{cjg+12}  &  \url{http://bootes.iaa.es/en/}  \\
Mt. Siding Spring, Australia & Skymapper & 31\degr 16\arcmin S & 149\degr 04\arcmin E & 1163 m  &  \citet{ksb+13}  &  \url{https://en.wikipedia.org/wiki/Siding_Spring_Observatory} \\
Sutherland, South Africa & MASTER-SAAO &  32\degr 17\arcmin S & 20\degr 18\arcmin E & 1760 m  &   \citet{master}  &   \url{http://observ.pereplet.ru}  \\
La Serena, Chile & DECam (CTIO) &    30\degr 10\arcmin S & 70\degr 48\arcmin W & 2207 m  &  \citet{amw+12}  &  \url{http://www.ctio.noao.edu/noao/}  \\
La Silla, Chile & TAROT-LaSilla &  29\degr 15\arcmin S & 70\degr 43\arcmin W & 2400 m  &  \citet{Tarot}  &  \url{http://tarot.obs-hp.fr/infos/}  \\
Cerro Paranal, Chile & VST &  24\degr 37\arcmin S & 70\degr 24\arcmin W & 2600 m  &  \citet{vst} &   \url{http://www.eso.org/public/teles-instr/}  \\
Salta, Argentina & TOROS &  24\degr 36\arcmin S & 67\degr 19\arcmin W & 4650 m  & \citet{toros}  &  \url{http://toros.phys.utb.edu/}  \\
Mauna Kea, Hawaii & W. M. Keck &   19\degr 49\arcmin N & 155\degr 28\arcmin W & 4145 m  &  \citet{keck}  &  \url{http://www.keckobservatory.org/about/} \\
Haleakala, Hawaii & PanSTARRS & 20\degr 42\arcmin N & 156\degr 15\arcmin W & 3052 m  &  \citet{panstarrs} &  \url{http://neo.jpl.nasa.gov/programs/} \\
Canary Islands, Spain & Liverpool & 28\degr 45\arcmin N & 17\degr 52\arcmin W & 2363 m  &  \citet{ssr+04} &  \url{http://telescope.livjm.ac.uk/About/} \\
Hanle, India & Hanle & 32\degr 47\arcmin N & 78\degr 52\arcmin E & 4500 m  & \cite{prabhu06} &   \url{http://www.iiap.res.in/iao_site}  \\
Palomar, USA & PTF & 33\degr 21\arcmin N  &  116\degr 51\arcmin W & 1712 m  &  \citet{lkd+09} &  \url{https://en.wikipedia.org/wiki/Palomar_Mountain} \\
Mt. Ontake, Japan & Kiso &   35\degr 47\arcmin N & 137\degr 37\arcmin E & 1130 m  &  \citet{tis+77}  &  \url{http://www.ioa.s.u-tokyo.ac.jp/kisohp/TELS/tels_e.html}  \\
\enddata
\tablecomments{See Figure~\ref{fig:obsloc} for a plot of these locations.}
\end{deluxetable}

\subsection{Simulated GW events}\label{subsec:simevent}
We use gravitational wave events from binary neutron star coalescence simulations by \citet{spf+14}, who used realistic detector sensitivity for LIGO--Hanford (H), LIGO--Livingston (L) and Virgo (V) at various stages of the gravitational wave network to recover the injected events that meet pre-defined detection thresholds. They calculate sky localization of these events using 
\sw{BAYESTAR}~\citep{sp16}, and supply the products as \sw{HEALPix} files~\citep{ghb+05}. \citet{spf+14} had simulated detections by LIGO in a period from 18 Aug to 19 Oct for two observing sessions, with gravitational wave detector sensitivity corresponding to the O1 and O2 observing runs. 
We note that the actual sensitivities attained in LIGO--Virgo observing runs may be somewhat different from their adopted values, thereby altering the localizations to some extent. 
The final data set\footnote{The simulated localization files are available at \url{http://www.ligo.org/scientists/first2years/}.} contains 630 two-detector events at O1 sensitivity; while for O2 sensitivity it has 365, 15 and 14 events for HL, HV and LV respectively, and 81 three-detector events with O2 sensitivity. 
Their dates do not match the actual dates of the O1 GW observing run, and are inconsistent with expected dates for O2. Further, the dates may introduce a seasonal bias in the comparison of various locations, as southern observatories will get longer nights in northern summer, and vice versa. Hence, we need to move these simulated events to different dates for comparison.

The sensitivity of LIGO, and hence the localization of detected events, is fixed in geocentric coordinates~\citep{fairhurst09}. The simulated GW detections can thus be reassigned to any other time when the relative orientations of the geocentric and celestial coordinate systems are the same. Thus, the event localization region in celestial coordinates remains unchanged if any event is moved to the same sidereal time on another day. As a further generalization, events can be moved to an arbitrary time stamp by considering the sky localization in geocentric coordinates, and transferring it to appropriate celestial coordinates at the new time stamp~\citep[see for example][]{eok+15}.

\subsection{Selection of dates}\label{subsec:dates}
To disentangle the effects of location and seasons, we first consider an idealized case where all detections are on the dates of the equinoxes, where all sites on earth have the same amount of night time. To evaluate the extent of seasonal variations, we also simulate observations for cases where all events occur at the summer or winter solstice. These cases are discussed in detail in \S\ref{sec:SESE}.

Next, in \S\ref{sec:fullsim} we consider a set of dates for the first and second LIGO--Virgo observing runs, O1 and O2. For O1, we use the actual dates: 18 Sep 2015\footnote{O1 start date: \url{www.ligo.caltech.edu/news/ligo20150918}.} to 12 Jan 2016\footnote{O1 end date: \url{www.ligo.caltech.edu/news/ligo20160112}.}. 
For O2, we consider specific possible dates to allow our analysis to be performed, and the likely
split into two parts. We consider O2A, with Hanford and Livingston detectors, to span the period from 1 Dec 2016 to 28 Feb 2017. For our example, Virgo is taken to join these two detectors in O2B, spanning the period from 1 Apr 2017 to 31 May 2017. Considering that not all detectors will be functioning throughout these phases, we undertake separate analysis of O2B into two--detector events (\S\ref{sec:O12-2Det}) and three--detector events (\S\ref{sec:3Det}).

\subsection{Analysis}\label{subsec:analysis}
We load and analyze the \sw{HEALPix} files in python using \sw{healpy} and \sw{astropy}~\citep{astropy}. We consider the 99\% credible region for GW localization, and hereafter refer to it as a ``patch''. Observations are simulated for a period of 24~hours from the trigger, and limited to night time (sun at least 18\degr\ below the horizon). We also impose a upper bound on the zenith angle of observations, based on two principles: most telescopes cannot point arbitrarily close to the horizon, and quality of data is poor for observations at high airmass\footnote{Airmass is the relative optical path length for light through the Earth's atmosphere, and is set to unity for a source exactly overhead.} (high zenith angle). We choose an airmass of $\sim2.5$, so that only parts of the patch that rise at least 24\degr\ above the horizon are observed. We do not include any constraints based on lunar phase or lunar angle.

After filtering out the \sw{HEALPix} pixels satisfying these conditions, we sort them by probability of containing the EM counterpart, and add up the probability for the top $N$~\sqd. To enable comparisons with real telescopes, we have considered different cases with $N$ = 1, 3, 10, 30, 100 and 300~\sqd 

For reference, we also consider a best-case scenario that could be attained by say a space telescope, free of horizon and twilight constraints. In this case, we only mask out pixels within a 42\degr\ sun avoidance angle\footnote{As we have selected 18\degr\ twilight and a 24\degr\ altitude constraint, any observatory anywhere in the world will not be able to observe any point $<42\degr$ from the sun.}. This case is marked as ``best'' in the following sections.

We note two important caveats in these analyses: we have not considered the shape of a telescope field of view, or the duration of visibility of the patch. For instance, if a patch is irregularly shaped, each telescope image may include several parts of the sky that are outside the 99\% patch~\citep[cf.][]{gbn+15}. Thus, the telescope may image say 100~\sqd, but cover only 50~\sqd\ of our patch. This inefficiency is sensitive to the exact shape and size of the field of view. There might also be scenarios where a large part of the patch is visible from the site, but only for a short duration of time: for instance, an event occurring overhead just before morning twilight. However, the time taken to image $N$~\sqd\ to a given depth can vary drastically between different telescopes. This in turn may limit the fraction of the visible patch that can be imaged through the night. As our focus is to compare geographic locations rather than telescopes and instruments, we do not consider these two effects in our simulations.

\section{Geographical and seasonal effects at Solstices and Equinoxes}\label{sec:SESE}

\begin{figure}[!thbp]
\begin{center}
\includegraphics[width=0.49\textwidth]{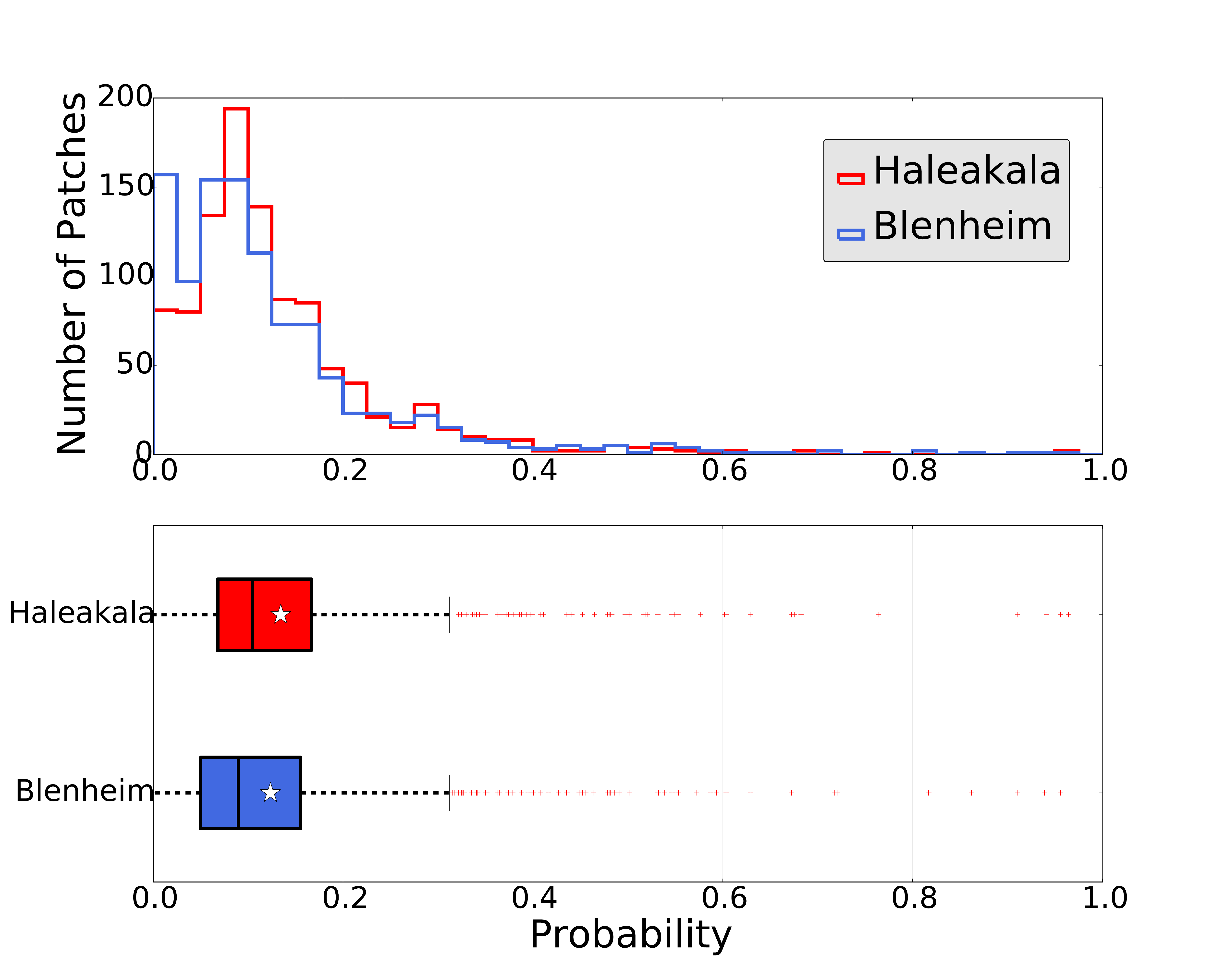}
\caption{Comparing the observable probabilities for Blenheim and Haleakala, sites of BOOTES3 and PanSTARRS respectively. We simulated observations of 1024 GW events detected by the two-detector network (O1 or O2), with all events moved to the date of the autumnal equinox. We assume that each site has a telescope capable of imaging 30~\sqd\ to the requisite sensitivity. The overall histograms are comparable. The slightly poorer observability from Blenheim results from a greater distance from the equator, which makes some northern patches completely inaccessible. \textit{Upper panel:} Histograms of observable probabilities for all events. \textit{Lower panel:} Box and whisker plots for the same histograms.}
\label{fig:comp-NS}
\end{center}
\end{figure}

As discussed in \S\ref{subsec:dates}, we first compare the various observatory locations in terms of their coverage for GW events on equinoxes. First, consider the case of all 1024 (O1 + O2) two-detector events moved to the Autumnal equinox. For each event, we calculate the probability observable from each location, and find that all sites have comparable performance. As an example, in Figure~\ref{fig:comp-NS} we show histograms of the probability of finding the counterpart by imaging the best visible 30~\sqd\ from two sites --- Blenheim, New Zealand and Haleakala, Hawaii --- which respectively had the worst and best median performance in this category. It is seen that both sites have a comparable performance.

In order to simplify visual comparisons, in the rest of this paper we use box-and-whisker plots (Figure~\ref{fig:comp-NS}, lower panel). The filled box spans the central 50\% of the histogram, extending from the lower quartile to upper quartile. In other words, 25\% of the events have observable probability less than the left edge of the box, while it is greater than the right edge for another 25\%. The range between these two points is called the inter-quartile range (IQR). ``Whiskers'' plotted on either side of the box extend to 1.5$\times$ IQR\footnote{If the histograms were Gaussian, the ends of the whiskers would be at  4.7$\sigma$ on either side of the mean.}. Any outlier points outside the whiskers are marked by red `+' signs. Since we are primarily interested in properties of the distribution rather than specific outliers, we have often scaled the plots such that some of the outliers are beyond the plot limits. The line and star inside the box show the median and mean of the distribution.

\begin{figure}[thbp]
\begin{center}
\includegraphics[width=0.49\textwidth]{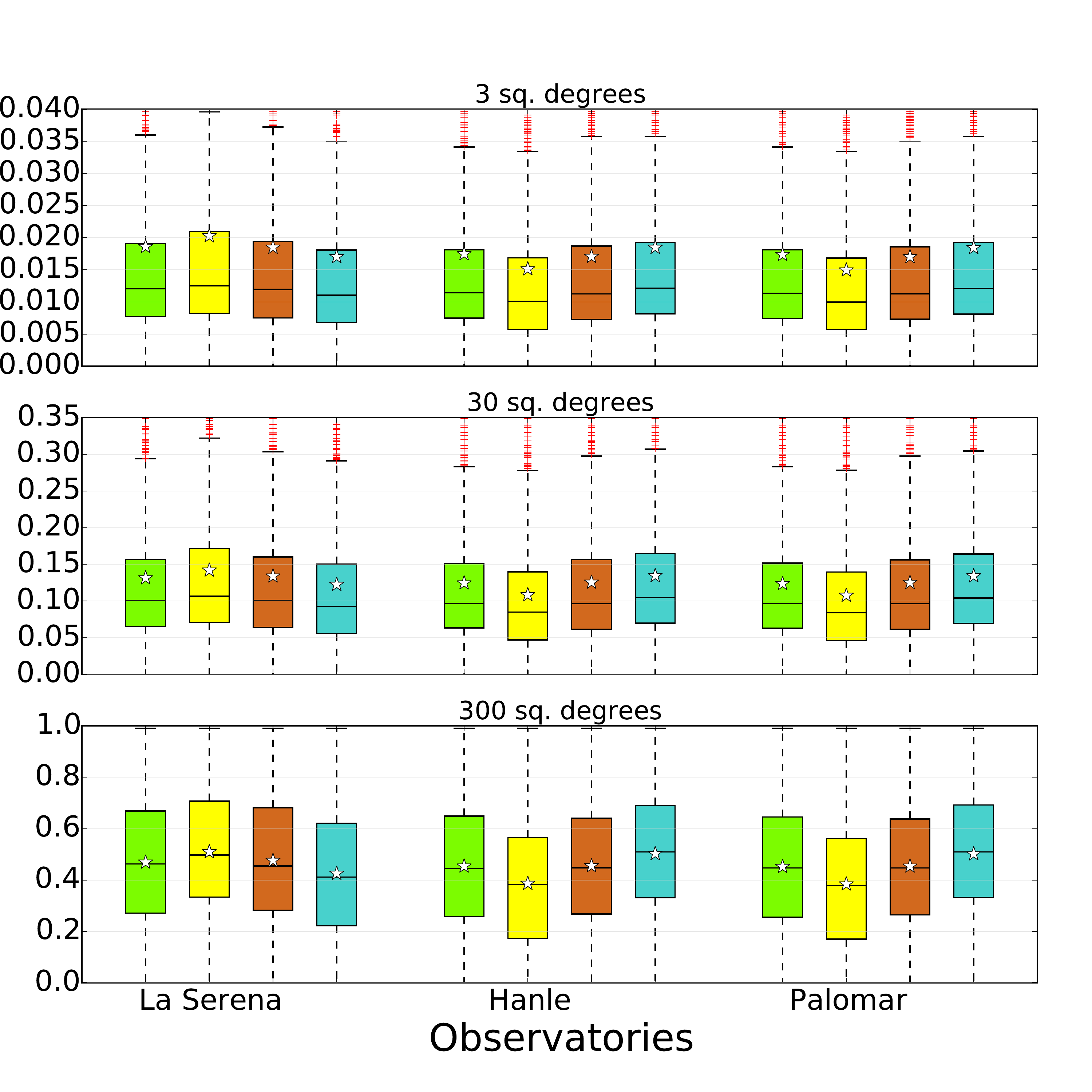}
\caption{Location-wise performance comparison between La Serena, Hanle, and Palomar, for all the O1 and O2 two-detector events. Events were shifted to the day of vernal equinox (green), summer solstice (yellow), autumn equinox (chocolate) and winter equinox (blue). The three panels distribution of observable probabilities are calculated using telescopes that can image 3, 30 and 300~\sqd\ respectively within 24~hours of the trigger.}
\label{fig:Equi-Sol-Box}
\end{center}
\end{figure}

Next, we investigate the effect of seasons by simulating observations of the same 1024 (O1 + O2) two-detector patches, but moved to the dates of the solstices. As expected, we see that northern observatories perform better during the winter solstice, owing to longer nights; while southern observatories perform better during the summer solstice. For example, Figure~\ref{fig:Equi-Sol-Box} shows the performance of La~Serena, Hanle, and Palomar~Mountain on the equinoxes and solstices, showing clearly the reversal of favored seasons between the northern and southern hemispheres.

\begin{figure*}[thbp]
\begin{center}
\includegraphics[width=\textwidth]{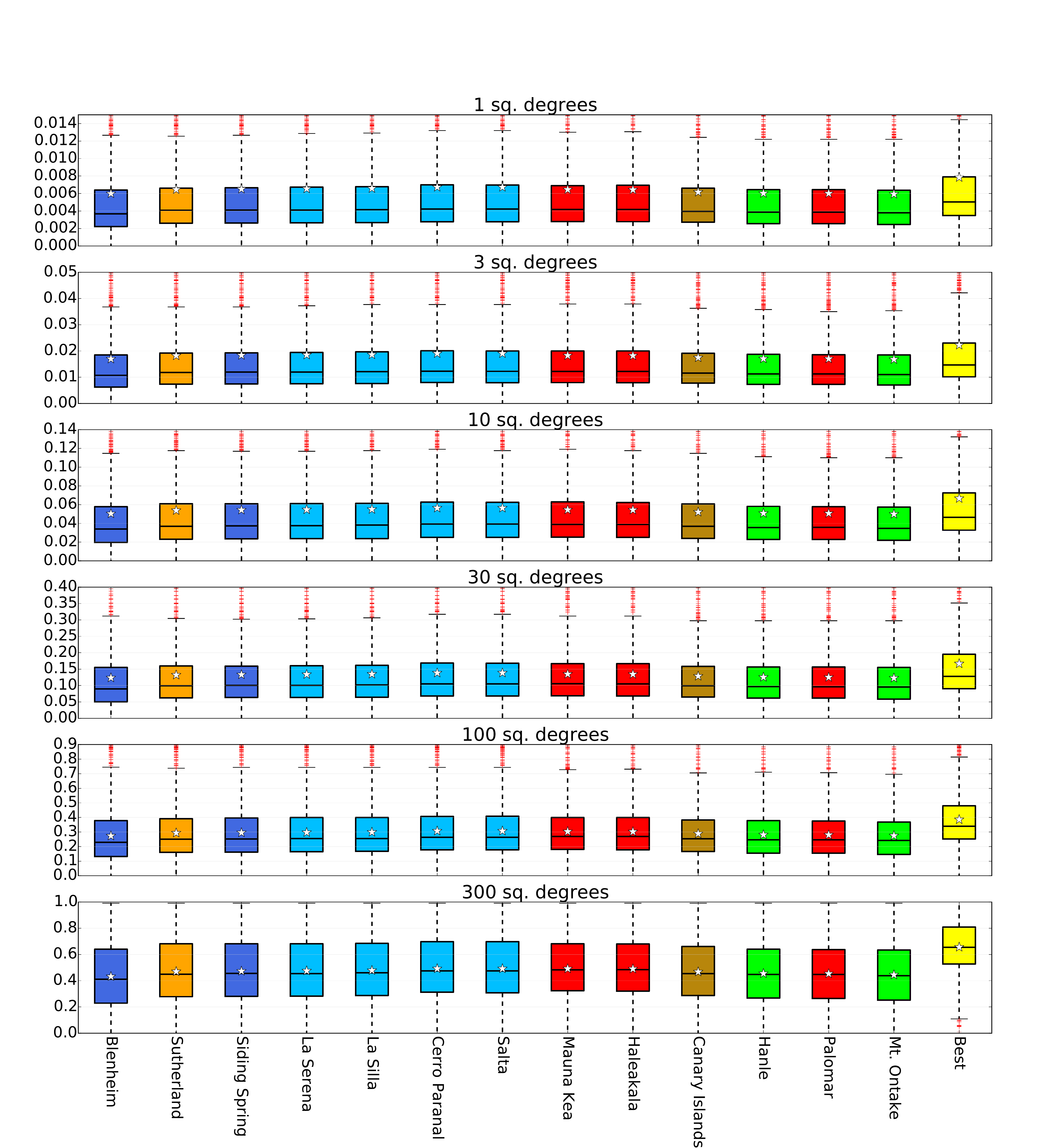}
\caption{Probability of finding optical counterparts for simulated two-detector events on the autumnal equinox. The simulation sample includes events with O1 and O2 sensitivity. The observatories are sorted by longitude, and color-coded by continent as in Figure~\ref{fig:obsloc}. The best-case scenario considering only solar exclusion angle but ignoring horizon constraints is plotted in the rightmost column. On comparing the location-wise performance for 1, 3, 10, 30, 100 and 300~\sqd, we see that all sites perform comparably with a very slight trend along the latitude.}
\label{fig:Equi-Box}
\end{center}
\end{figure*}

\begin{figure}[!thbp]
\begin{center}
\includegraphics[width=0.49\textwidth]{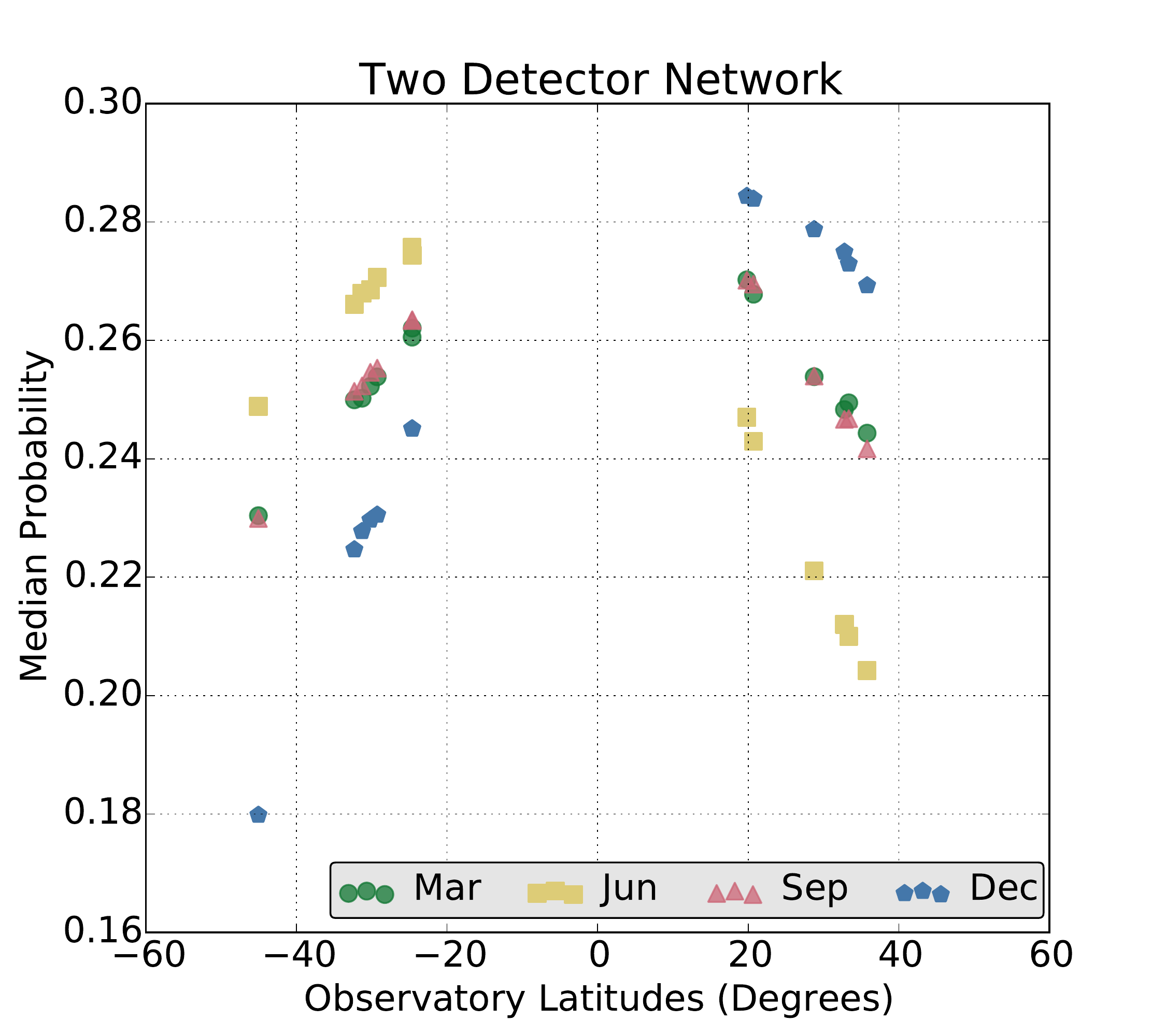}
\includegraphics[width=0.49\textwidth]{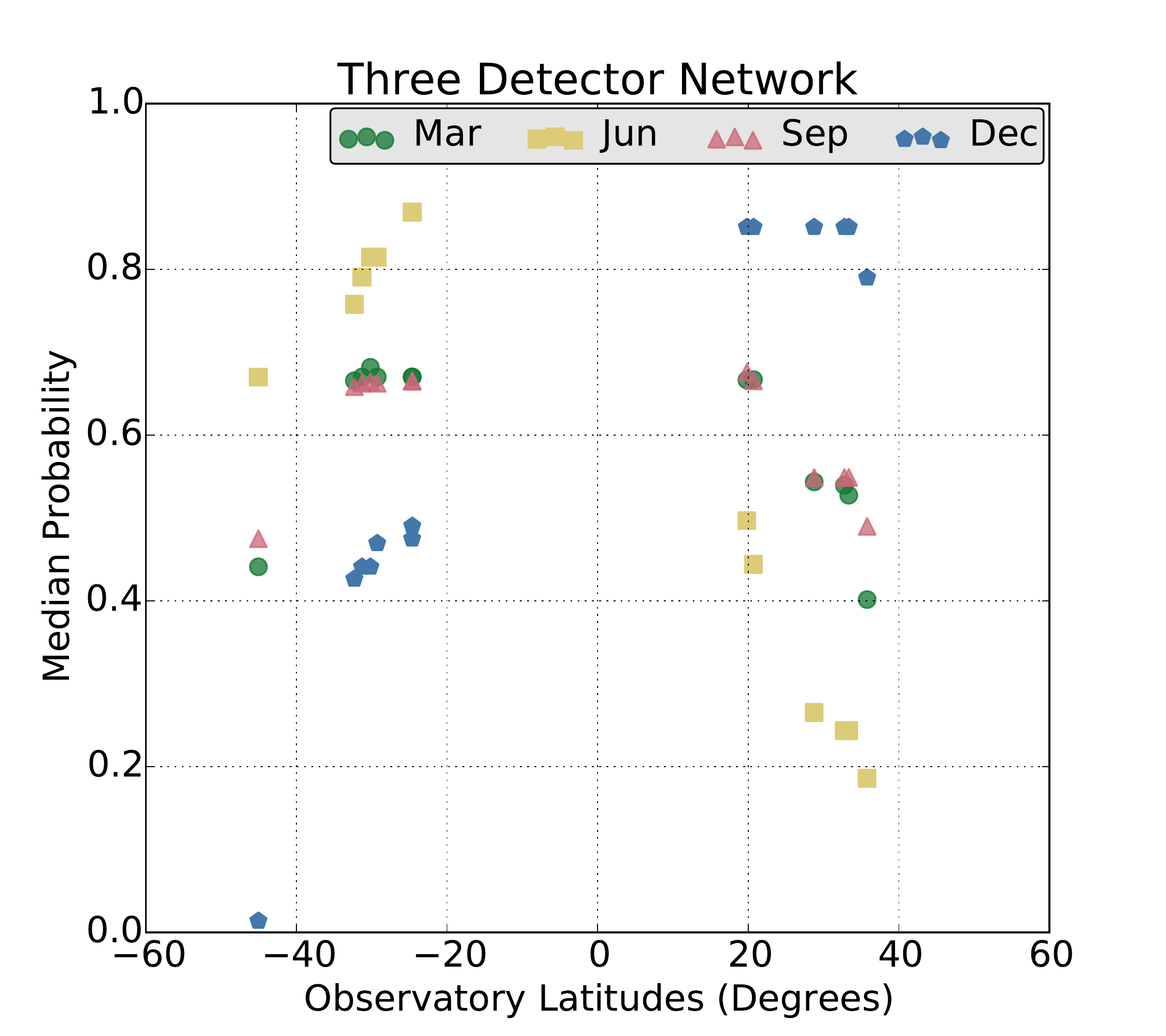}
\caption{Effects of seasons on median observable probability for different latitudes. The different colors and symbols show median observable probability on the days of the vernal equinox (green circles), summer solstice (yellow squares), autumnal equinox (pink triangles) and winter solstice (blue pentagons). \textit{Upper panel:} For each site, we compute the median of the probability covered for 1024 (O1 + O2) two detector events, with a telescope limited to observing 100~\sqd\ during the night. \textit{Lower panel:} Median probability of finding the counterpart for a source localized by a the three detector HLV network, with the same telescopes limited to observing 100~\sqd\ during the night.
}
\label{fig:Equi-Sol-Sum}
\end{center}
\end{figure}

While the overall performance of the various sites for equinox observations is similar (Figure~\ref{fig:Equi-Box}), looking at the median values of observable probabilities shows some interesting trends (Figure~\ref{fig:Equi-Sol-Sum}). We see that on the equinoxes, sites at mid-latitudes have a few percent higher probability of finding the optical counterpart of a gravitational wave event, as compared to observatories in the temperate zones. This can be explained by a combination of two effects: (i) the two LIGO detectors detect more sources at mid-declinations as compared to equatorial or polar declinations, and (ii) sites further from the equator have a progressively smaller fraction of the sky accessible even on the equinox night. Similar effects have also been discussed in \citet{cev+16}.

We also see that if we cover 100~\sqd\ of the sky from any given location, the median observable probability for events on the day of the solstice changes by 1\% to 7\% with respect to equinox, and these differences are stronger for sites further away from the equator (Figure~\ref{fig:Equi-Sol-Sum}, upper panel). This seasonal variation stems primarily from the duration of the night, determining the fraction of the sky visible. The effect is limited to a few percent due to the large areas and long arc-like shapes of GW events localized by just two detectors~\citep{spf+14}. One would then expect the seasonal differences to be more stark if the localization improves. In the limiting case, if GW sources were pinpointed on the sky by the gravitational wave detectors, the observable probability would be governed by the latitudinal variation of detector sensitivity function \citep{fairhurst11} and would vary more strongly with the fraction of the sky visible at night. Indeed, this is the case with improved localizations provided by a network of three gravitational wave detectors. We repeat our simulations using GW events that were detected by the HLV network, and find that solstice-to-solstice changes in the median observable probability can be as high as 60\% (Figure~\ref{fig:Equi-Sol-Sum}, lower panel).
For reasons discussed in \S\ref{sec:3Det}, results for the three-detector network should only be considered qualitatively, not quantitatively.

\section{Example observing runs}\label{sec:fullsim}
As a specific case, we repeat our simulations using the actual dates of the first LIGO science run (O1) and a set of example dates of O2. Although the dates of O2 are uncertain, we aim to give an overall perspective of how observatories at different locations may perform under these conditions. The qualitative nature of these results will be insensitive to $\sim$10 day shifts in dates.

\subsection{Two Detector Network}\label{sec:O12-2Det}
Our two-detector sample consists of 630 events with O1 sensitivity, and 394 events at O2 sensitivity. The latter are split as 365, 15 and 14 events for HL, HV and LV respectively. The relatively smaller number of Virgo-detected events arises from expected sensitivity and uptime of the three detectors, as discussed in \citet{spf+14}. As discussed in \S\ref{subsec:dates}, O2 is expected to be subdivided into O2A and O2B. For our simulations, we distribute the 365 HL events randomly in both parts of the run, and keep the 29 HV/LV events in O2B.

\begin{figure*}[thbp]
\begin{center}
\includegraphics[width=\textwidth]{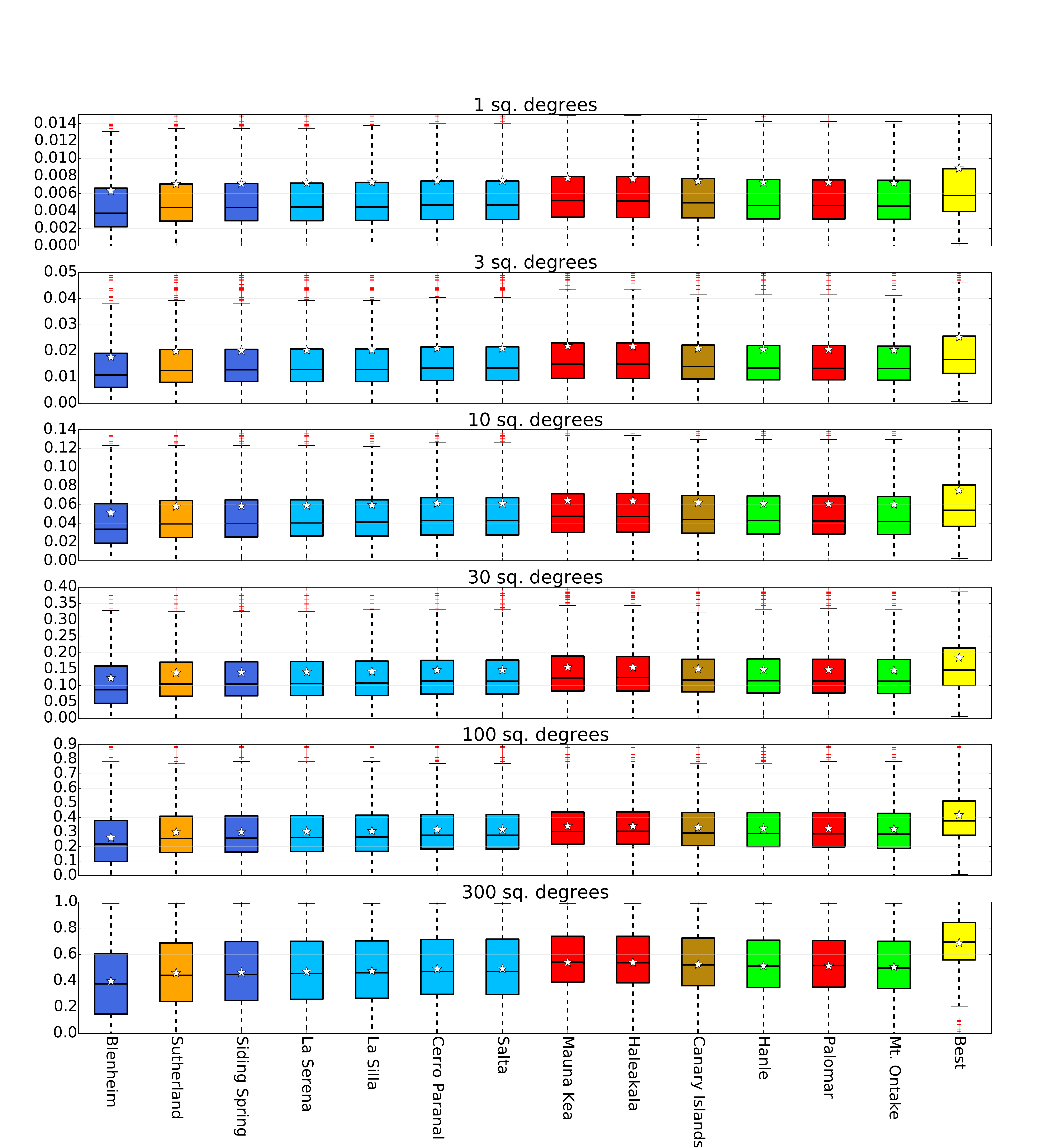}
\caption{Comparisons of various observatories for O1. 630 simulated events detected with Hanford and Livingston GW detectors at O1 sensitivity were randomly distributed over actual dates of O1. The box-plots of observable probabilities are as in Figure~\ref{fig:Equi-Box}. Northern observatories had better chances of finding EM counterparts as compared to southern ones.}
\label{fig:BP-O1-2Det}
\end{center}
\end{figure*}

\begin{figure*}[thbp]
\begin{center}
\includegraphics[width=\textwidth]{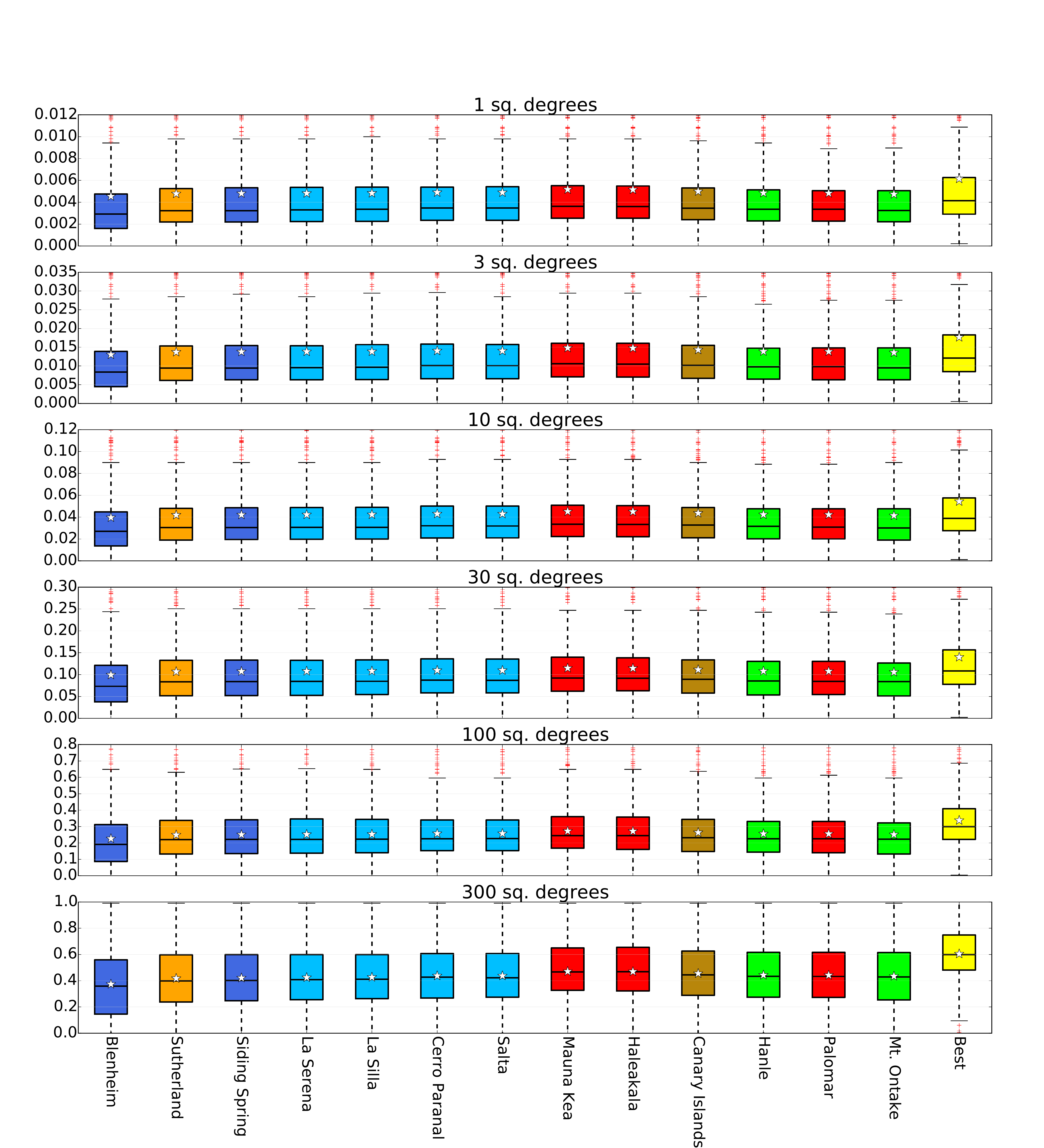}
\caption{Comparisons of various observatories for O2. 394 simulated events detected with GW detectors at O2 sensitivity were distributed over example dates of O2 as described in \S\ref{sec:3Det}. The box-plots of observable probabilities are as in Figure~\ref{fig:Equi-Box}. Operational periods O2A and O2B favor locations in different hemispheres, as a result, all observatories have comparable odds of finding EM counterparts of GW sources.}
\label{fig:BP-O2-2Det}
\end{center}
\end{figure*}

As O1 was conducted during northern winter, one expects northern observatories to perform better than southern ones, and this expectation is borne out by simulations (Figure~\ref{fig:BP-O1-2Det}). 
Mauna Kea and Haleakala have the best chance of discovering an optical counterpart, with a median probability of 0.30 for a camera capable of imaging 100~\sqd\ in a night. Blenheim, the southernmost location in this study, had a median probability of 0.22 of imaging the optical counterpart with the same resources. Incidentally, the localization of GW150914 happened to peak in the southern skies~\citep{aaa+16}, while localizations of GW151226 and LVT151012 were more uniformly spread over declination~\citep{lsc16}. Thus, small number statistics worked in favor of southern observatories in O1.

The assumed split dates of O2 span approximately northern winter and spring, slightly favoring northern observatories in O2A and southern ones in O2B. The net result is that the timing of observing runs slightly favors northern observatories, but the overall performance of observatories is dominated by their latitude, following a similar trend as the equinoxes (Figure~\ref{fig:BP-O2-2Det}). The number of two-detector detections involving Virgo in O2B is rather small, and does not alter the trends in any significant manner.

\subsection{Three Detector Network: HLV}\label{sec:3Det}
The joint detection of any gravitational wave event by all three GW detectors drastically changes the follow-up scenario. The median area encompassing 90\% probability of containing the true source drops from several hundreds of square degrees to few tens of square degrees~\citep{spf+14}. We now investigate how this affects the follow-up from various locations.

\begin{figure}[!bhtp]
\begin{center}
\includegraphics[width=0.49\textwidth]{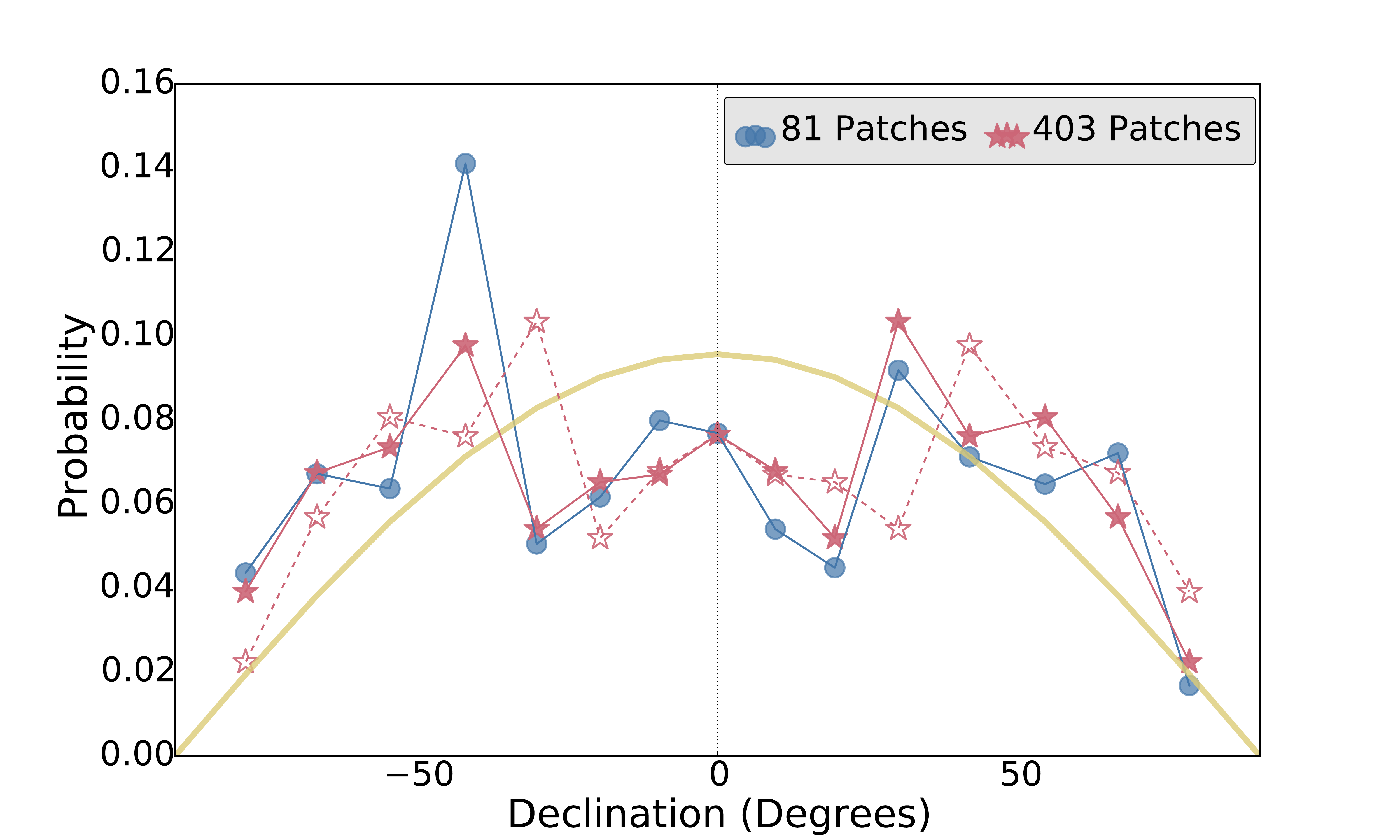}
\caption{Probability of source location as a function of declination, averaged over multiple patches. The solid yellow curve is proportional to $\cos(\delta)$, the expected probability distribution if the sensitivity of gravitational wave detectors was independent of direction. Blue circles show the distribution for the average of the original 81 patches from \citet{spf+14}, with a strong peak at $\delta \sim -40\degr$. Solid brown stars show the average probability in each declination bin for the average of the final 403 patches used in this work. To guide the eye, hollow brown stars connected with dashed lines show a mirror image of the final probability distribution: showing that the final set has reasonable north-south symmetry.}
\label{fig:B4After}
\end{center}
\end{figure}

\citet{spf+14} provide only 81 events detected by the HLV network. As the net area of each localization patch is a small fraction of the entire sky, any study using this sample will suffer from small number statistics. Indeed, averaging the all-sky probabilities for these 81 events shows a bias towards the southern skies (Figure~\ref{fig:B4After}, blue circles). We work around the this problem in two steps. First, we translate these patches in trigger times (and right ascensions) to get 400 new sky localizations as discussed in \S\ref{subsec:simevent}. Next, we randomly select 20\% of the patches with localization peaking in the southern hemisphere, and drop them from the set. This nearly removes the unexpected north-south asymmetry in the full sample (Figure~\ref{fig:B4After}, brown stars). This gives us a final set of 403 patches, which we distribute randomly over the dates of O2B for simulating followup of three-detector events.

\begin{figure}[thbp]
\begin{center}
\includegraphics[width=0.49\textwidth]{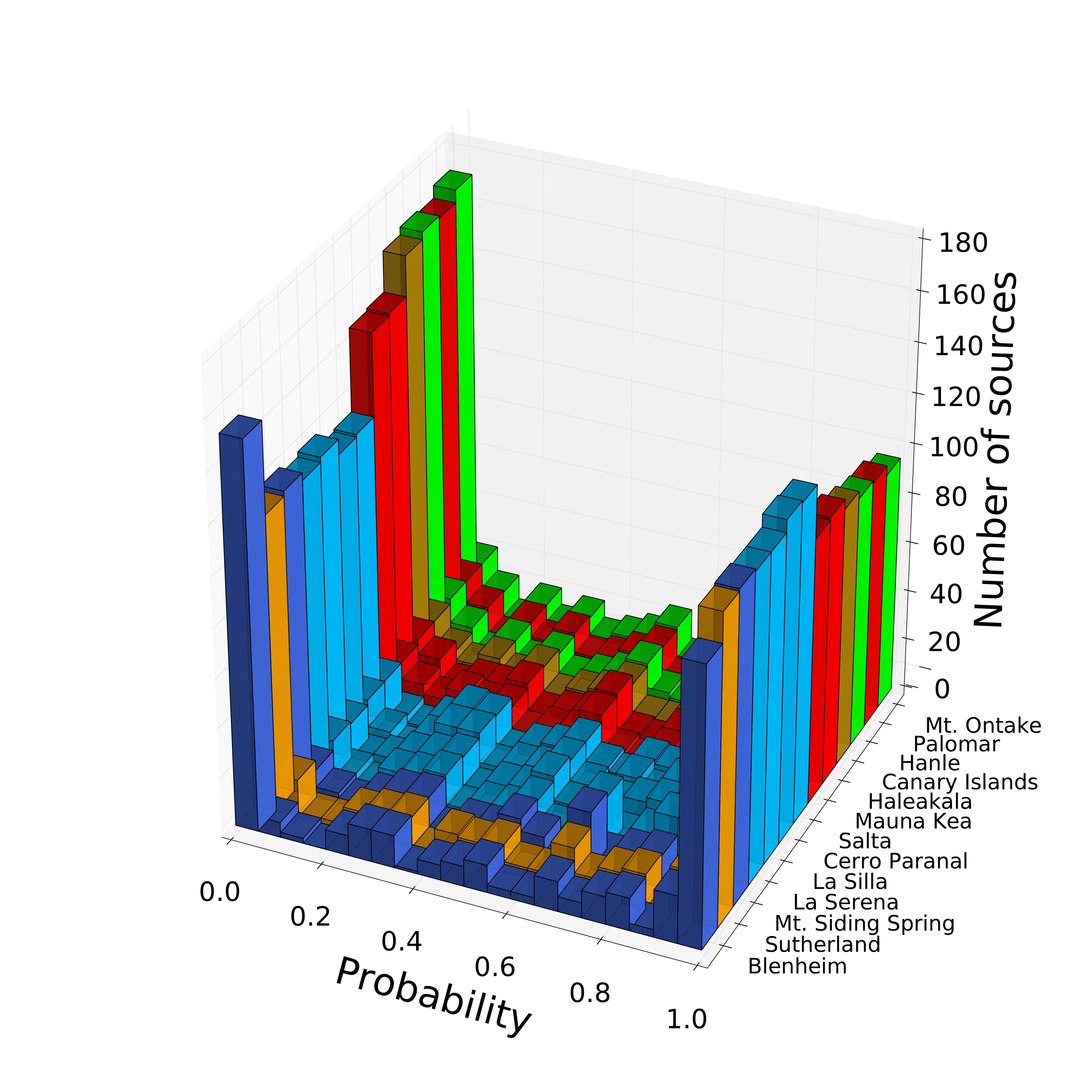}
\caption{Probability of imaging the counterpart of a gravitational wave event from various locations for the 403 events recovered with a three gravitational wave detector network. We have assumed that instruments at each location can observe 30~\sqd\ of the sky. As many events have rather small 99\% localization areas, observations will be dominated by whether the localization region is visible at all. A large fraction of events have $p_{\rm obs} < 0.05$ or $p_{\rm obs} > 0.95$.}
\label{fig:3Det-30}
\end{center}
\end{figure}

\begin{figure*}[thbp]
\begin{center}
\includegraphics[width=\textwidth]{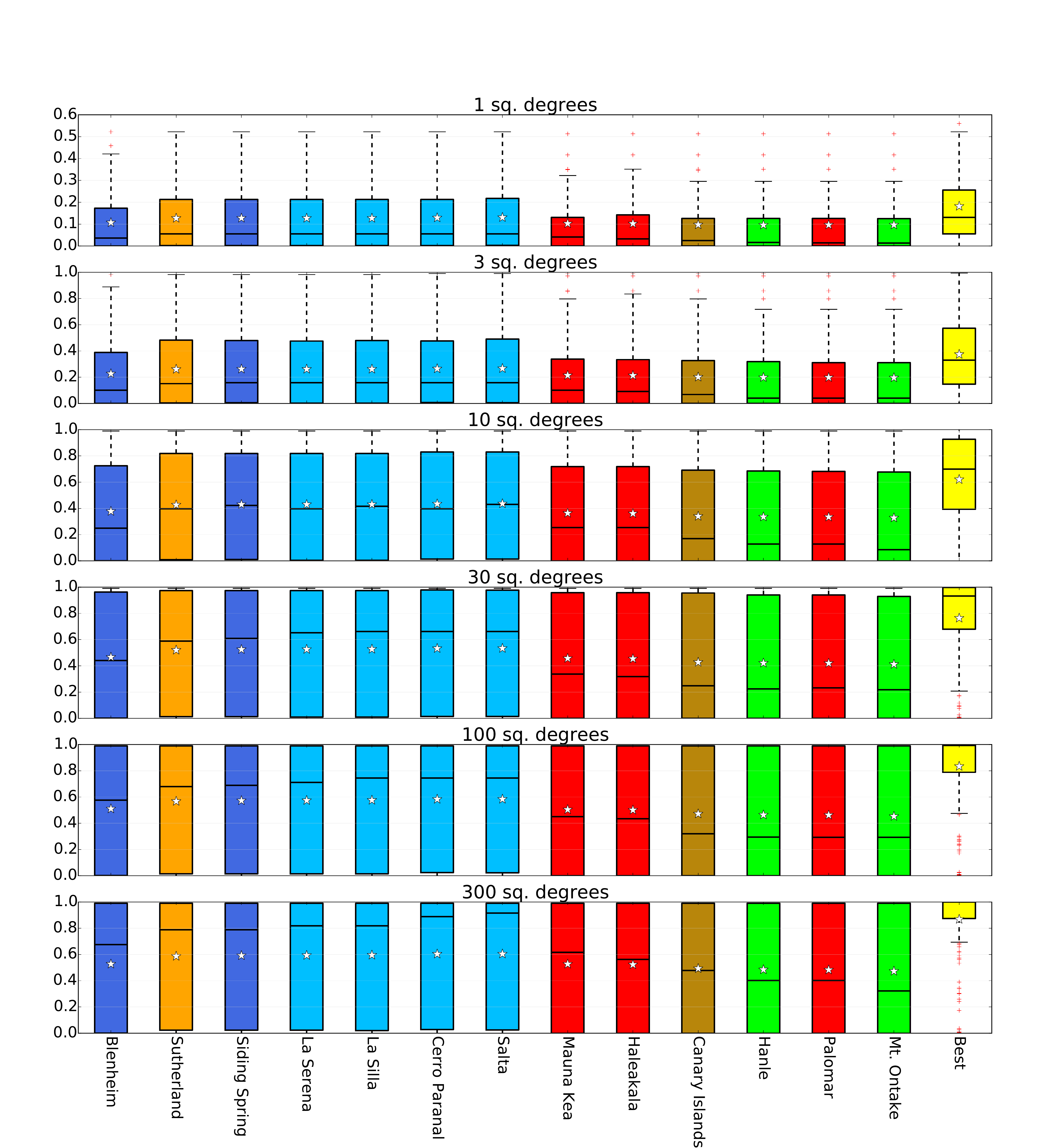}
\caption{Probability of finding optical counterparts for simulated three-detector events with example O2B dates. Details are as in Figure~\ref{fig:Equi-Box}. While box-plots are not the best representations of these bi-peaked distributions (see Figure~\ref{fig:3Det-30}), we use them for consistency with other plots. The better performance of southern locations can be attributed primarily to the season. As we increase the sky coverage, the mean probability of finding a counterpart increases rather slowly beyond 30\degr.}
\label{fig:3Det-Box}
\end{center}
\end{figure*}

The well-constrained sky localizations from the three detector network lead to very different distributions of observable probability as compared to a two detector network. Observatories at most locations can cover nearly the entire patch if it rises at that location, but cover almost zero probability otherwise. Figure~\ref{fig:3Det-30} highlights this effect for telescopes that can cover 30~\sqd\ in a single night. At all observatories, a large fraction of events have $p_{\rm obs} < 0.05$ or $p_{\rm obs} > 0.95$. The overall result is that the observable probability from any given location (Figure~\ref{fig:3Det-Box}) is completely dominated by seasonal effects. For the dates of O2B used in this work, our simulations show the southern locations stand a better chance of finding optical counterparts of gravitational wave sources. As the 403 patches used in these simulations were generated from just 81 events, we caution the reader that these results should be interpreted qualitatively.

\section{Discussion}\label{sec:discussion}
We investigate the effects of observatory locations on the probability of discovering optical/infrared counterparts of gravitational wave sources. We show that the odds of discovering EM counterparts show some latitude dependence, but weak or no longitudinal dependence. 

Seasons have a much larger effect on the observability of GW localization regions, and dominate over geographic variations. These effects too are stronger for observatories at high latitudes, where the length of the night is most strongly affected by seasons. \citet{cev+16} have independently reached similar conclusions by using a different methodology and slightly different assumptions.

Our simulations show that northern observatories had slightly better odds of discovering the EM counterparts of GW sources in O1, though the small number statistics of just two detections and one candidate dominated over this effect. Based on our assumed sample dates of the second observing run O2, all observatory locations have comparable chances of finding EM counterparts. 

In mid-2017, O2B may discover the first three-detector GW event, with much better localization than two-detector networks: simplifying ground-based follow-up. This season favors southern observatories, giving them significantly higher chances of discovering counterparts of such GW events.

In order to compare the performance of various optical/infrared observatory sites, we have assumed identical equipment at all locations. In practice, this is not the case and instrument characteristics like imaging depth and field of view will play a strong role in successful detection of an EM counterpart. Coordinated observations among multiple sites~\citep{sps12}, use of galaxy catalogs~\citep{hmv14,gck+15} and enhanced scheduling algorithms~\citep{chm+15,gbn+15,rsg+16} will help observatories to boost their chances of detecting electromagnetic counterparts.

We have taken a representative sample of observatories and considered a set of example dates of LIGO--Virgo observing runs for our simulations. To facilitate further exploration on these lines, our simulation codes are available at \url{https://github.com/emvarun/followup-and-location}. Users will also have to download the \citet{spf+14} data set from \url{http://www.ligo.org/scientists/first2years/}. In these python codes, users can easily add/remove sites and change simulation dates. The codes produce a set of plots, summary tables, and a detailed table for the observable probability of each event from each location.

\section*{Acknowledgments}

We thank P.~Ajith, Hsin-Yu Chen, Reed Essick, Shaon Ghosh, N.~K.~Johnson-McDaniel and Leo Singer for helpful discussions.

VB acknowledges the financial support of Department of Science and Technology, Government of India for the ``Global Relay of Observatories Watching Transients Happen (GROWTH)'' project which prompted this study. GROWTH is a part of ``Partnerships for International Research and Education (PIRE)'', jointly funded by NSF grant No. 1545949. APR acknowledges the ICTS-S.~N.~Bhatt Memorial Excellence Fellowship Program for his summer at ICTS during which a large initial part of the work was done. AG's research was supported by the AIRBUS Group Corporate Foundation Chair in ``Mathematics of Complex Systems'' at ICTS, and by the Max Plank Society and the Department of Science and Technology, Government  of India, through a Max Planck Partner Group at ICTS. SB acknowledges support from NSF grant PHY-1506497.

This LIGO document P1600278-v2.

\software{NumPy~\citep{numpy} and Matplotlib~\citep{matplotlib}, Astropy \citep[\url{http://www.astropy.org}]{astropy}, 
  HEALPix~\citep{ghb+05}}, Healpy (\url{https://healpy.readthedocs.org/}), Ephem (\url{https://pypi.python.org/pypi/pyephem/}).

\bibliographystyle{aasjournal}
\bibliography{geographic}
\end{document}